# Notes on Pervasive Virtuality

Luis Valente[1], Bruno Feijó[2], Alexandre Ribeiro Silva[3], Esteban Clua[1]

[1]MediaLab, Institute of Computing, UFF, Brazil

[2]VisionLab, Department of Informatics, PUC-Rio, Brazil

[3]Instituto Federal do Triângulo Mineiro, Brazil

{lvalente,esteban}@ic.uff.br, bfeijo@inf.puc-rio.br, alexandre@iftm.edu.br

**Abstract.** This paper summarizes notes about a new mixed-reality paradigm that we named as "pervasive virtuality". This paradigm has emerged recently in industry and academia through different initiatives. In this paper we intend to explore this new area by proposing a set of features that we identified as important or helpful to realize pervasive virtuality in games and entertainment applications.

## 1 Introduction

An important goal in virtual reality applications is to immerse the user's senses in an artificial virtual environment (VE) through an interactive experience. A key factor regarding how this interactive immersive experience is successful refers to the sense of presence (Sanchez-Vives and Slater 2005). Experimenting with virtual reality is becoming a hot topic in industry and academia as several device manufactures have recently started to bring affordable HMD hardware to the consumer market. Some of these experiences resulted in projects that explore a common pattern in a mixed-reality environment for entertainment: players wear mobile HMD devices (seeing only virtual content) and are able to move freely in a physical environment, being able to touch physical walls and interact with physical objects while immersed in the simulation. In this paper we present notes about "pervasive virtuality", which is term we coined to denote concepts and the environment where this type of entertainment application takes place and present a set of features that we identified as important to realize this new paradigm.

## 2 Pervasive virtuality

Pervasive virtuality comprises a mixed-reality environment that is constructed and enriched using real-world information sources. This new type of mixed reality can be achieved through the use of non-see-through HMD devices, wireless networking, and context-aware devices (e.g. sensors and wearable technology). In PV, the user walks through a virtual environment by actually walking in a physical environment (exposed to sounds, heat, humidity, and other environmental conditions). In this virtual environment, the user can touch, grasp, carry, move, and collide with physical objects. However he/she can only see virtual representations of these objects. Even when a user is physically shaking hands with another user, he/she has no idea about the real characteristics of this other user (e.g. as gender, appearance, physical characteristics).

This new type of mixed-reality application emerged recently, with several similar initiatives appearing almost simultaneously in the industry ("The VOID" (The VOID 2016) and "real virtuality" (Artanim 2016)) and academia ("live-action virtual reality game" (Silva

*et al.* 2015)). An earlier similar academic initiative was "virtual holodeck" (Steinicke *et al.* 2008). The experience in this new type of mixed-reality is extremely intense and immersive. However, being a new area, the literature lacks proper definitions, design principles, and methods to guide designers and researchers. This paper contributes to shed more light on these issues. We start by defining PV considering the traditional reality-virtuality continuum (Milgram and Kishino 1994).

## 2.1 Comparison with traditional reality-virtuality continuum

Milgram and Kishino (1994) proposed a mixed-reality taxonomy with visual displays in mind. Yet, most of the mixed-reality applications found in the literature simply juxtaposes real and virtual objects through the projection of visual artefacts. For instance, a common example of augmented virtuality is the video of a real human face projected on a 3D mesh of an avatar's head in a virtual world. Essentially, in these applications, "augmented virtuality" consists of a virtual world augmented with the mapping of an image or video from the real world in virtual objects, and "augmented reality" is the same process the other way round. Therefore, we need to extend Milgram and Kishino's taxonomy to accommodate other forms of mixed reality.

We need a taxonomy that can cope with situations where real, physical, objects are transformed into virtual objects, and vice-versa (i.e. virtual objects become real objects). We propose to identify the first situation as "pervasive virtuality" and the later situation as "ubiquitous virtuality". These situations represent a better fusion of reality and virtuality, which goes beyond a simple mapped visual projection. In these new environments, transformed objects should work like a proxy. Figure 1 illustrates a proposal to extend the Milgram and Kishino's taxonomy with these new concepts.

Augmented virtuality is different than pervasive virtuality, as in the first one real world objects are projected in virtual content and the HMDs that the user wears (if any) necessarily are see-through devices. Examples of augmented virtuality can be found in (Bruder *et al.* 2010; Paul *et al.* 2005).

Pervasive virtuality is different from the mixed-reality found in "pervasive games" (Valente *et al.* 2015b), although they may share similarities (e.g. context-awareness). Essentially, pervasive games are based on the idea of a real-world augmented with virtual content (left of Figure 1) through context-awareness, made possible through sensor devices placed in the physical environment and carried by players (in mobile and wearable devices) while they move in the real-world.

Our concept of ubiquitous virtuality (UV) is aligned with well-known definitions of this term in the literature (Kim *et al.* 2006), which means that this type of mixed-reality integrates virtual objects seamlessly in the real environment and preserves as many human senses as possible. However, transforming virtual objects into real ones is a much more complicated affair and we have almost no examples to produce. Computational holography is a potential technology in this regard, but this is a research area still in its infancy. Smart materials that can change their shape in particular magnetic fields or change their texture in response to a voltage change could also be used, but they are in an experimental stage. In the realm of science fiction, UV would be the Star Trek Holodeck. Figure 1 illustrates that PV enhances visual virtual environments towards what we consider the true "virtual reality" – an environment with a strong sensation of immersion and presence due to the existence of several human senses. On the other hand, UV tends to evolve towards what we call "real virtuality" – a virtual world so convincing that we cannot easily distinguish it from the real world. As a last observation about Figure 1, we can expect that a continuum of mixed reality situations connect its two ends (Real Virtuality and Virtual Reality).

PV transforms physical objects into virtual equivalents (also named "proxy objects" in (Steinicke *et al.* 2008)) by representing their geometry in the virtual environment and tracking them with wireless networking systems. Furthermore, pervasive virtuality maps the real environment into the virtual environment through the compression and gain factors that apply to a user during his/her immersion in virtual environments. For example, the user's real movements always occur over smaller (and curved) paths when compared to his/her virtual paths.

## 2.2 Defining and characterizing PV

In the context of the previous section, we define pervasive virtuality (PV) as follows. PV is a mixed reality where real, physical, objects are transformed into virtual objects by using real-world information sources through direct physical contact and context-aware devices (e.g. sensors and wearable technology). In PV users wear non-see-through HMDs all the time, which means that they do not see real-world content. PV may require intensive use of compression and gain factors on external world variables to adjust the transformation between reality and virtuality. PV can be better understood by exploring its characteristics, as Section 3 describes. This section summarizes the concepts underlying those characteristics.

All content that a user experiences in PV is virtual – the simulation uses digital content and generates virtual content based on real world information sources. These information sources are: (1) the physical environment architecture; (2) physical objects that reside in the environment; and (3) context information (see Sections 3.1 and 3.6).

PV takes place in a "simulation stage" (or "game stage", in the case of games), which consists of a physical environment (e.g. a room, school floor, or museum) equipped with infrastructure to support the activities (e.g. wireless networking, sensors, and physical objects). The simulation (or game) uses these elements to create the mixed reality.

In PV, a user wears a non-see-through HMD device and walks in the physical environment (Section 3.3), being able to touch physical walls and other elements. The user sees a 3D virtual world through the HMD and does not see the real world. The simulation constructs a virtual world based on the physical environment architecture (i.e. the first information source, as a 1:1 matching), keeping these two worlds superimposed (Sections 3.1, 3.7). The simulation detects physical objects (e.g. furniture, portable objects, and users' bodies – the second information source) and maps them into virtual representations, which are then displayed to the user (Section 3.1). Users touch, grasp, carry, and move these physical objects, but they only see their virtual representation (Sections 3.1, 3.5).

The third information source is context information (Dey 2001) (Section 3.6), which the simulation may use to generate virtual content and to change the rules or simulation behavior (i.e. unpredictable game experiences and emergent gameplay). Examples of context information include: 1) player information (e.g. physiological state, personal preferences, personality traits); 2) physical environment conditions (e.g. temperature, humidity, lighting conditions, weather); 3) information derived from the ways a player interacts with physical objects and input devices; and 4) information derived from relationships and interactions among players in the game (the social context).

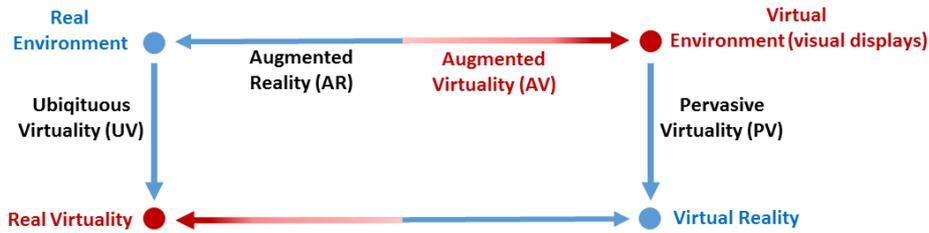

**Fig. 1** Extending Milgram and Kishino's taxonomy of real and virtual environments

A PV application may respond back to the user through various channels and various types of media (Section 3.4). Some of these channels may be worn or carried by users (Section 3.1), and some of them correspond to physical objects that are spread in the physical space (e.g. smart objects and environment devices, Sections 3.4, 3.6). Finally, users may interact in PV through multiple modalities (e.g., voice, body movements, and gestures), ordinary physical objects, and context-aware devices, supporting tangible, body-based, and context-aware interaction paradigms (Section 3.5).

## 3 Pervasive virtuality features

Table 1 lists the initial set of pervasive virtuality features in two levels. On the first level there are seven features (or aspects), which are subdivided in a second-level that represent more specific aspects. The remaining of this section describes these features.

### 3.1 Virtuality

*Virtuality (Vir)* regards handling the virtual aspects in PVs, including *Virtual world generation (VWG)* and *Virtual content presentation (VCP)*.

#### 3.1.1 Virtual world generation

*Virtual world generation* regards procedures to generate the 3D virtual world based on physical world structure, which may be a real-time process or may be a pre-configuration step (i.e. mapping the environment before the simulation runs), for example.

#### 3.1.2 Virtual world presentation

*Virtual content presentation* concerns issues about how the simulation presents the virtual world and virtual content to users. For example, PV may present content through HMDs, wearable devices (capable of providing haptic feedback), the underlying physical structure, and physical objects (e.g., touching physical walls, tables, and holding small objects). In case of HMDs, there are issues that the simulation must address properly, such as adverse effects on users (e.g. nausea, motion sickness). Hearing may be stimulated by isolating or non-isolating headphones and smart objects (Section 3.4).

### 3.2 Sociality

*Sociality (Soc)* refers to social aspects and social implications of the simulation. Currently we identify *Social presence* and *Ethical concerns* as sub-aspects.

#### 3.2.1 Social presence

*Social presence (SP)* concerns *"how people experience their interactions with others and refers to conditions that should be met in order to experience a sense of co-presence (i.e. mutual awareness)"* (Wolbert *et al.* 2014). Social presence may happen among real people and/or among users and virtual characters. For example, a simulation may foster social relations by providing team play activities that require collaboration. A stronger possibility in this ex-

ii

ample are activities that require collaboration due to complimentary user roles.

### 3.2.2 Ethical concerns

*Ethical concerns (Eth)* regards topics related to the well-being of users and ethical issues. For example, Madary and Metzinger (2016) discuss issues about virtual environments influencing user's psychological states while using the system, lasting psychological effects after the simulation is over (e.g. long-term immersion effects, lasting effects of the illusion of embodiment, undesired behaviour change).

## 3.3 Spatiality

*Spatiality (Spa)* regards aspects related to the physical space usage. Currently we identified *Mobility* as a sub-aspect.

### 3.3.1 Mobility

*Mobility* concerns issues related to the free movement of players in the physical environment. Some examples are providing adequate network support that covers the entire simulation stage, requirements of user movement due to physical space size and interaction with physical objects, and how to simulate a large virtual environment in a confined space.

## 3.4 Communicability

*Communicability (Comm)* concerns aspects about the infrastructure that the simulation uses to communicate with users and other simulation components

### 3.4.1 Connectivity

*Connectivity (Con)* refers to the networking infrastructure that is required to support activities in the simulation (and associated issues). For example, PVs may require wireless local networking with specific requirements (e.g. low latency).

### 3.4.2 Game stage communication

*Game stage communication (GSC)* refers to the communication channels that the game uses to exchange information with players in the game stage. For example, environment devices (Valente *et al.* 2015b) are objects placed in the physical environment and may output information (e.g. audio), generate effects in the physical world (e.g. smells, wind, heat, cold, spray water, open doors, move elevators).

## 3.5 Interaction

*Interaction* (Int) refers to interaction paradigms in pervasive reality. So far we identify these important paradigms that contribute to create the "live-action play" aspect of PV: *Tangible interaction (TI)*, *Body-based interaction (BI)*, *Multimodal interaction (MI)*, and *Sensor-based interaction (SI)*. These interaction paradigms may be facilitated through wearable devices (e.g. smart bands, motion sensors) and dedicated infrastructure (e.g. motion-capture cameras).

*Multimodal interaction (MI)* corresponds to interaction through multiple modalities such as voice input, audio, and gestures. *Tangible interaction (TI)* represents the "tangible object metaphor" as first defined by Ishii and Ullmer (1997). In TI, a user interacts with the simulation by manipulating context-aware mobile devices (e.g. portable device equipped with sensors) and ordinary physical objects (e.g., wood sticks, rocks). In PV, the simulation tracks all these devices and display a correspondent virtual representation to the user through a HMD. For example, a real wood stick could become a virtual sword in the simulation. *Body-based interactions (BI)* represents interactions through body movements such as jumping, spinning, walking, running, and gestures. *Sensor-based interaction (SI)* represents interactions of implicit nature (Rogers and Muller 2006) based on sensor devices.

| First level | Second level |
|---|---|
| Virtuality (Vir) | Virtual content presentation (VCP), Virtual world generation (VWG) |
| Interaction (Int) | Tangible interaction (TI), Body-based interaction (BI), Multimodal interaction (MI), Sensor-based interaction (SI) |
| Communicability (Comm) | Connectivity (Con), Game stage communication (GSC) |
| Spatiality (Spa) | Mobility (Mob) |
| Sociality (Soc) | Social presence (SP), Ethical concerns (Eth) |
| Resilience (Res) | Uncertainty handling policy (UHP), Game activity pacing (GAP), Mixed-reality consistency (MRC) |
| Context-awareness (CA) | Dynamic content generation (DCG), Gameplay adaptability (GpA) |

**Table 1. Pervasive virtuality features**

An example is "proximity interaction", where the simulation triggers events when a sensor detects the presence of a user.

## 3.6 Context-awareness

*Context-awareness (CA)* refers to acquiring and using context information in simulation activities. This is a key component of pervasive reality. Dey (2001) defines context as *"any information that can be used to characterize the situation of an entity. An entity is a person, place, or object that is considered relevant to the interaction between a player and an application, including the user and applications themselves"*. The simulation is able to sense context information through several means, such as: 1) The physical place where the game happens may host physical objects that are equipped with sensors (e.g. smart objects in (Valente *et al.* 2015b)). These devices may be connected to other similar objects and to game servers; 2) The player carries or wears devices that sense context from the environment and/or from the player; and 3) The game queries remote information about the player based on the player identity (e.g. social network profiles).

### 3.6.1 Gameplay adaptability

A system can be considered as "context-aware" if *"it uses context to provide relevant information and/or services to the user, where relevancy depends on the user's task"*. In this regard, "context-awareness" means that pervasive reality is able to adapt the gameplay according to the current context conditions (*Gameplay adaptation – GpA*).

### 3.6.2 Dynamic content generation

*Dynamic content generation (DCG)* refers to the capacity of creating virtual content dynamically based on context information. When this feature is taken to the extreme, the simulation generates all virtual content dynamically. For example, in the context of pervasive games, *Insectopia* (Peitz *et al.* 2007) generates important game content (i.e. insects) based on Bluetooth device ids that the game scans while the players are wandering around.

## 3.7 Resilience

*Resilience (Res)* refers to how a game is able to cope with technology uncertainties (i.e. in sensors and networking) to prevent them from breaking immersion and the simulation experience. These un-certainties stem from inherent technology component limitations such as accuracy, precision, response time, and dependability. For example, in PV the tracking problem is a key issue as: 1) sensor technologies ignore physical world boundaries; 2) sensor technologies may not be able to track objects moving above a certain speed threshold. We identify three important aspects for resilience: *Uncertainty handling policy (UHP)*, *Game activity pacing (GAP)*, and *Mixed-reality consistency (MRC)*.

### 3.7.1 Uncertainty handling policy

*Uncertainty handling policy (UHP)* refers to specific strategies a simulation may use to handle uncertainties. Valente et al. (2015b) discussed five general strategies to approach these issues (e.g. hide, remove, manage, reveal, exploit).

### 3.7.2 Game activity pacing

*Game activity pacing (GAP)* refers to how the pacing of activities might interfere with the operations of specific technologies (e.g. sensors), and vice-versa, and what to do about this issue.

### 3.7.3 Mixed-reality consistency

*Mixed-reality consistency (MRC)* refers to how a game keeps the physical and virtual worlds superimposed and synchronized (in real-time) without negative side-effects that a player might perceive. For example, a key element in the PV pipeline is tracking of physical objects and physical elements (e.g. architecture). PV may use different approaches, such as sensors or computer vision. A PV simulation can process tracking through the mobile user equipment (e.g. "inside processing" in portable HMDs, cameras, and computers) or through a dedicated infrastructure (e.g. "outside processing" using cameras fixed in the physical infrastructure and dedicated computers).

## 4 Conclusions

This paper summarized notes about a new mixed-reality paradigm that has emerged due to recent advances in HMD technology. This is a work in progress and it complements our previous work (Valente *et al.* 2015a) where we started to explore this area.

We refrained from trying to come up with concise definitions about this new paradigm and decided to pursue a path similar to the one Valente et al. (2015b) underwent: discover important features (or qualities) that contribute to the uniqueness of a specific type of game. In case of Valente et al. (2015b), the specific type of game paradigm is pervasive games. We believe that this approach may lead to more practical results (e.g. helping designing and developing these applications) instead of trying to formulate abstract definitions – this is an approach that echoes other works about fuzzy definitions in game research such as "pervasiveness" (Nieuwdorp 2007). The initial feature set is not a complete set, but it is a starting point to foster discussion and further research agendas.

## Acknowledgements

The authors thank CAPES, FINEP, CNPq, FAPERJ, and NVIDIA for the financial support to this paper.